# Did the concepts of space and time change that much with the 1905 theory of relativity?

Mario Bacelar Valente

Abstract

The 1095 theory of relativity is rightly considered as a breakthrough moment in the history of physics; in particular. it is widely accepted that it brought a new conception of space and time. The purpose of this work is to reevaluate to what point and in what sense can we consider that the conception of space and time went through a transformation when going from Newtonian mechanics to the theory of relativity.

1 Introduction

In this work I will consider the theory of relativity conception of space and time from two perspectives: (1) an historical one; that of the development of the theory of relativity in the context of the existent Lorentz's electron theory (that entails a different conception of space and time); (2) an an-historical one, in which one compares different space-times associated with, what is usually considered as more and more fundamental theories/approaches; in this case I will compare the space-times that are compatible with Newtonian mechanics with the space-time of the theory of relativity. The first is the approach of the historian (see, e.g., Janssen 1995); the second is, usually, the approach of the philosopher (see, e.g., Friedman 1983).  Along the different sections of the work I will be jumping in between these two methodologies, basically making my own blending of them.
     Regarding the concept of space, on one side I will look at it from the perspective of Einstein's development of a view in terms of an homogeneous vacuum (without an ether that enables to have a sort of absolute positions), and to which one needs to 'add' a reference frame; on the other side I will consider space in Newtonian mechanics in terms of the manifold approach and Newton's original views, and see how a relativist Galilean space without the notion of absolute space can be brought into the theory. I will arrive at a point of 'junction' in which we see that both Newtonian mechanics and the theory of relativity can be seen as sharing a common notion of space that, in the lack of a better word, might be called spatiality. The part dedicated to space can be found in the sections 2 to 4.
     In what regards time, to which are dedicated in particular sections 5 to 9, I will make a reevaluation of the conception of time in the theory of relativity. Like in the case of the so-called length contraction in relation to the notion of space, the so-called time dilatation does not have to be seen as implying by itself a throughout revision of the concept of time. There is nothing new about this view; Einstein himself took the time dilatation to be a "purely kinematical consequence of the theory" (Einstein 1911, 348). As Einstein mentions, what seems to be important to a newer conception of time is the so-called relativity of simultaneity, which he also refers to as the relativity of time. Regarding this point I will follow two different lines of approach. Dieks (1998, 2006) proposes a view in which the central aspect of a renewed conception of time in the theory of relativity can be found in the concept of proper time. Dieks says that we experience time locally. In this view the relativity of simultaneity might even be seen as not relevant for the conception of time. Dieks arrives at the view that there are different now-points associated to each worldline. In one line of approach I will present a view different from Dieks's in which actually there is only one now, and this now should not be thought in terms of a temporal concept; instead it should be seen as a spatial concept: the spatial co-existence of physical systems and processes is what we call now (present). What Dieks and I agree on is on the centrality of the concept of proper time. In the second line of approach I will reevaluate the so-called relativity of simultaneity in terms of this view of now as spatial co-existence. I will defend that the so-called relativity of simultaneity does not bear

on the notions of past, present and future since it is in fact a kinematical effect like the length contraction and the time dilatation. We can see that simply by noticing that the so-called relativity of simultaneity is in fact a relativity (i.e. dependence on the reference frame) of synchronization of distant clocks.

There still seems to be quite a difference between the Newtonian conception of time and the one brought by the theory of relativity. However one can complete the 'transformation' of the Newtonian space-time into a truly relativist Galilean space-time, not only by letting go of absolute space but by freeing the Galilean time from Newton's absolute time, and instead adopting a relativistic proper time view also for the case of the Galilean space-time (I take Brown's approach to be doing this implicitly; see Brown 2005, 19-20). With this last move one arrives at a conception of time that has much in common with the one set forward in the theory of relativity.

2 Newtonian and Galilean space and time

If one addresses Newtonian space and time adopting as a starting elements the notions of manifold[1] and of coordinate system[2] one soon faces a strange situation. Mathematically one describes space and time as a four-dimensional manifold. If one now considers a particular worldline (defined with the help of a coordinate system) – which we can regard, for example, as the set of spatial-temporal events[3] associated with a material physical system (e.g. an astronaut moving outside a space ship) –, for each temporal moment there is a tri-dimensional spatial plane and all the spatial points of this plane are simultaneous with the physical system's time, i.e. we have an absolute time. This implies that the four-dimensional space-time is 'stratified' in a temporal succession of instantaneous tri-dimensional spatial planes exhausting the manifold (see, e.g. Friedman 1983, 71-4; Earman 1989, 28). If we associate an 'intuitive' notion of space with this tri-dimensional planes, we end up with a space concept that is derived from more basic notions (manifold and coordinate system) that pops in and out of 'existence' at each temporal moment. To retain some spatial 'continuity', one is (almost) obliged to provide a relation between the initially unrelated points of the different planes by imposing an additional structure to the manifold: a rigging of non-intersecting geodesics filling the manifold so that one can say that two point from different planes corresponds to the same position. We arrive then at an absolute space.

When compared with Newton's original formulation, one sees that Newton started by considering time as independent (flowing on its own, independently of any physical systems), regular (the self-time-intervals are identical, even if there is no 'independent' way of verifying this), and mathematically described (see, e.g., Brown 2005, 18-9).[4] Space is for Newton mathematically

---

1   A manifold is a very general mathematical structure that enables the definition of the different mathematical space and time one presently encounter in physics: the Newtonian or Galilean space-time; The Einsteinian or Minkowski space-time; the curved space-time of Einstein's gravitation theory (see, e.g., Friedman 1983).
2   Mathematically a coordinate system is a map of an open subset of an n-dimensional manifold into an open subset of $\mathbb{R}^n$ (see, e.g., Wald 1984, 11-3). Physically a coordinate system is defined within a reference frame, which is seen as an actual complex material system consisting of quasi-rigid objects serving as measuring rods and physical systems that empirically are shown to go through regular phases at the same pace serving as clocks. An actual material reference frame gives operational procedures to measure spatial distances and temporal durations, which are the basis to define a coordinate system within the reference frame (see, e.g., Reichenbach 1927, 14-24 and 113-9; see also Bohm 1965, 42-7; Norton 1993, 835-6; Darrigol 2007). Within an analysis of the theory of relativity one usually idealizes a reference frame as a cubic latticework or grid of measuring rods with a clock at every intersection (see, e.g., Taylor and Wheeler 1965, 17-9).
3   An event is idealized as a time measurement made by a clock located at a particular point in a reference frame, in this way giving also a distance measurement; it is associated with a particular physical process (like a thunder striking just next to the clock) or a physical system (like a particle moving within the reference frame and passing by the clock).
4   According to Torretti the mathematical description of time can be traced back at least to Aristotle in his treatment of Zeno's paradox (Torretti 2007, 67-80). The immediate predecessor of Newton in the geometrical description of time is Galileo that in his study of motion took time as another geometrical dimension in an equivalent way to space (see, e.g., Clavelin, 1968). This 'spatialization' of time, in which time is mathematically describe as a sort of spatial described, independent and previous to matter, and absolute, i.e. there is a meaningful notion of self-position (see, e.g., Weyl 1949, 99). Newton recognizes that only relative spaces (a notion in some aspects similar to the modern notion of reference frame) can be determined, i.e. that only relative positions can be measured. However by postulating that the center of the universe is at rest in relation to space, when measuring distances to the fixed reference one can determine absolute positions (see, e.g., Weyl 1949, 100).[5]

A crucial difference between Newton original formulation and the manifold approach[6] is that with Newton, space is a overwhelming presence previous to all entities and independent of the independently flowing time (that 'affects' matter but not space); on the other hand the manifold formulation seems to relegate space to a derived 'aspect' of a more fundamental structure (the manifold); it is dependent on a worldline defined in relation to a coordinate system (that is, on a material reference frame), turning out to be an almost virtual manifestation (instantaneous to a temporal moment). Also, as we have seen, in the second case the absolute nature of space must be built into the manifold by filling it with worldliness, which need matter to be meaningfully defined: the trajectories defining the rigging are the ones taken to be at rest (Friedman 1983, 78). This last aspect is, as we have seen, similar to Newton postulating a particular material system to be at absolute rest. The difference is that since Newton's absolute space is ever-present, a sole material system serves to relate relative positions to absolute ones. In the second case one needs to fill-up the manifold with worldlines; this would imply in Newton's perspective to fill all space with material things.

The selection of a particular rigging can be seen as the selection of a particular inertial frame that is actually at absolute rest regarding the absolute space (Friedman 1983, 87).[7] Returning to Newton's original formulation this reference frame is at rest in relation to the center of the universe, i.e. it is at absolute rest; in this way, this reference frame is always in the same absolute localization in relation to the underlying space. However, since Newton's mathematical description of motion does not depend on absolute positions, all inertial reference frames are equivalent, independently of being at rest or no. This led several authors to criticize Newton's notion of absolute space (see, e.g., Jammer 1993, 138-43). In fact Newton's theory is not crucially dependent on the notion of absolute space. This opens the door for a revision of Newton's notion of space by letting go of the idea of absolute position, i.e. the idea of a self-position of space itself. We arrive at what has been called Galilean space (see, e.g., Friedman 1983, 87; Weyl 1921, 149-159).[8]

From the manifold perspective this seems to be a simple procedure. One simply lets go of the rigging. One still has a distinction between accelerated and non-accelerated motion, but not the notions of absolute rest and absolute velocity (Friedman 1983, 87). The problem is that now we do not have a structure linking together more closely the instantaneous Euclidean planes that are, in the manifold approach, space.

From a Newtonian perspective things seem to be not better off. Without a self-position one loses a matter independent idea of position: space is still the underlying container but the spatiality

---

    dimension, is one of the defining moments of physics (see, e.g., Saint-Ours 2008).

5  Newton's approach presupposes a meaningful notion of rest in relation to space. A related idea is presupposed in Lorentz's electron theory in which one talks about the state of (absolute) motion or rest in relation to the ether (see below).

6  Regarding Élie Cartan's development of a covariant formulation of Newton's gravitation (which is based in a formulation of space and time in terms of a manifold structure) and latter developments see, e.g., Havas (1964); Misner, Thorne and Wheeler (1973, 289-301).

7  In an inertial reference frame, physical systems that are (approximately) free of interactions are at rest or move with rectilinear uniform velocity (i.e. are in relative rest or motion in relation to the reference frame). This definition is made in terms of Newton's first law. Another procedure is to define a reference frame in terms of one in which Newton's second law is (approximately) valid; thus an inertial reference frame is one in which there are no pseudo-forces, i.e. all the accelerations being measured correspond to actual interactions (see, e.g., DiSalle 2009).

8  Regarding the notion of absolute time it is not dependent on the notion of absolute space. Even if there are no preferred reference frames associated with 'absolute rest', the different inertial reference frames make an equivalent stratification of the manifold into instantaneous spatial planes momentarily simultaneous (see, e.g., Torretti 1996, 28).

between things, manifested in measurable distances, is only revealed with the things, i.e. without the notion of self-position (or equivalently of absolute space) there is no previous-to-things notion of position, measurable distance, and rest; these become matter-bond notions. The problem is mainly in the notion of measurable distance. The idea is that if I am here and the tree is over there, 10 m from me, I associate this measured distance to a spatial distance; it is a defining moment for a notion of space. The position is derived from this spatiality between the tree and me, as also is the case with the notions of rest and motion. If I forget myself and things and arrive at a notion of pure spatiality (i.e. space) independent and previous to all the rest, how can one maintain an yet to be 'condensate' distance (independent of a space-scale, i.e. the adopted distance unit) intrinsic to space without intrinsic space positions?

Independently of working in a Newtonian-like approach or a manifold-like one, when giving up on an absolute space (i.e. a notion of spatiality with intrinsic positions; or in other words a space with an intrinsic metric) there is the need for a new concept to anchor the establishment of distances and positions (i.e. to help defining a metric). This is made possible by resort to the key concept of reference frame. Inevitably this brings into the arguments material systems. We give up on the notion of self-position of space itself and consider a space filled with stable material things (that can be our rods and clocks) that enable setting forward a Galilean space-time. In other words a Galilean space-time needs material things to be an operationally meaningful metrical space-time. This change from an absolute space to a space where position (and rest and motion) are relative notions has as an identifying characteristic the notion of reference frame, which becomes a central element in setting forward notions of space (in the case of the Galilean space-time) or space and time (in the case of Einstein's space-time) compatible with relativity principles.[9] Contrary to the case of the Newtonian space where one can abstract away everything but a pure spatiality with its intrinsic metric, that is not possible in the Galilean space where besides a 'pure spatiality' metrically undefined, one needs this space to be filled with material objects to complete its metrical definition.[10]

3 Einsteinian space and time: a first look

Einstein sets forward is argumentation for a newer conception of space and time 'against' Lorentz's electron theory. In Lorentz's theory there is a material medium – the ether filling the whole of space. The ether is considered to be completely immobile in relation to space (see, e.g., Lorentz 1916, 11). What exactly this means is not really clear, since there is no way in which we might measure the ether's immobility. It must be taken as a defining sentence (similar to Newton's idea that the center of the universe is at rest) that more that giving us a valuable departing point to address the state of motion (or in this case of rest) of the ether, serves to pinpoint to the motion of things in relation to the ether; and so, one speaks not about the state of motion of the ether but about the state of absolute rest or absolute motion in relation to the ether. One may ask: what exactly is the underlying space in the ether theory? Is it a Newtonian absolute space, a Galilean 'relative' space or it can be either?

---

9  In my view, the 'advantage' of a Newtonian-like approach in relation to the manifold approach is that at least one still has a notion of spatiality (even if still without a metric) that is always present; not like in the case of the manifold where space seems to the an 'epiphenomena' of it, popping in an out of existence at each instant of the apparently more important and stable notion of absolute time, which, in this particular case, 'glues' all the instantaneous spaces into the manifold structure. This is a general trend of all manifold approaches to space and time. As Earman called the attention to "in a modern, pure field-theoretic physics, [the manifold] functions as the basis substance, that is, the basic object of prediction" (Earman 1989, 155).

10 Historically there was not a conceptual change from a Newtonian to a Galilean space-time involving sustaining the renewed conception of non-absolute space in terms of reference frame (i.e. material systems). As mentioned, there was a questioning of the notion of absolute space, but there was not an explicit construction of a Galilean space-time based on the central concept of reference frame; The change happened between the Newtonian absolute space (or, what is in practice equivalent to it: an ether filled space where it is apparently meaningful to define rest in relation to the ether) and Einstein's space-time.

The Newtonian space has already its self-positions in relation to which a material body might be in absolute rest. The ether, being supposed to be a sort of all-permeating matter, would be in identical circumstances with all other material things in what regards its state of motion in relation to the absolute space; if the ether is at rest in relation to the absolute space then all the material things that are at rest in relation to the absolute space are as well at rest in relation to the ether. The state of motion of material things in relation to the ether would derive/result from their common link to absolute space, i.e. it would be a relative state due to their identical state of rest in relation to the absolute space. When adopting a Galilean space without self-positions, if one tries to address the question of the state of motion of the ether in a Galilean space one faces the general difficulty mentioned above. Without spatially separated material things it seems difficult to posit an intrinsic metrical spatiality; one might try to consider this as a characteristic/property of the ether itself. The ether as a material continuum would 'contain' the spatiality usually attributed to space.[11] However, when adopting the more usual view of ether and space as two different concepts, it remains the fact that since in principle in a Galilean space there is no notion of absolute rest or motion of material things, the ether being material also does not have any absolute state of motion or rest in relation to space. What we might try to say, as it was done within Lorentz's electron theory, is that the material ether, since it pervades all things and fills space, provides a ever-present reference for other material things; in this case the state of motion or rest of things in relation to the ether would not be absolute in the previous sense of Newton's absolute space. It is in relation to this space that the notions of absolute rest or motion were developed; one might try to maintain the terminology in relation to the state of motion in relation to the ether but it has not the same meaning. In the first case it is a relation between matter and space; in the second case in is a relation between material entities. In any case, at least in practice, in the use of Lorentz's electron theory the self-positions of Newtonian absolute space would enter from the back door now as positions in relation to the ever-present ether (see, e.g., Einstein 1915, 248-50; Darrigol 2006).

In 1905, in his criticism of Lorentz's electron theory Einstein, like others (see, e.g., Darrigol 2000, 366-72), defended the view that we must "give up [on] the ether" (Einstein 1910, 124). Einstein argument is based on the idea that the ether seems to enable a special reference frame in relation to which things might still be said to be in 'absolute' rest or motion. Since it turns out that it is not possible to determine experimentally the velocity of material things in relation to the ether, there is no way to distinguish the ether's reference frame from other inertial reference frames. Accordingly, Einstein considers that one should "give up the notion of a medium filling all of space" (Einstein 1910, 124). Importantly, Einstein retains an idea of space that he will again and again characterized as a homogeneous vacuum (see, e.g., Einstein 1907, 258; Einstein 1910, 130; Einstein 1914, 4). The homogeneous space is a central element in the development of Einstein's newer views on space and time; in fact, *Einstein's construction of the 'physical time' depends on this previous already present spatiality*.

The argumentative line leading to the relativistic notions of space and time is based like in the case of Galilean space on the notion of reference frame. One starts not with a sort of pure spatiality which is not possible anymore since we do not count on self-position of space (i.e. an intrinsic metric) but by playing with two key ideas: (1) a spatial vacuum taken to be homogeneous (i.e. without any differentiating aspect of its own);[12] (2) different material things that share or adhere to this primeval spatiality. The (relatively) stable material things enable the possibility of material reference frames. It is within this spatial vacuum plus reference frame that the newer

---

11 Historically, a sort of 'coalescence' between the concepts of ether and space was in fact considered; for example, previous to the theory of relativity, Drude took the ether to be 'just' space with some physical properties (see, e.g. Kosto 2000, 18). Later, within his theory of gravitation, Einstein put forward the idea that the curved space-time is a sort of ether (Einstein 1920).

12 In my view one aspect of the homogeneity of space is that of being isotropic (i.e. there are no differentiating intrinsic directions in space). Usually homogeneity and isotropy are taken to be different assumptions (see, e.g., Cresser 2003, 13).

conception related to space and time are set forward.

In Einstein's view the most important aspect of the theory of relativity regarding the concept of time can be found in the so-called relativity of simultaneity of distant events (Einstein 1914, 4), to which Einstein also refers to as the relativity of time (Einstein 1915, 254). According to Einstein "two events that are simultaneous when observed from some particular coordinate system can no longer be considered simultaneous when observed from a system that is moving relative to that system" (Einstein 1905, 145).

To address the question of distant simultaneity one must first consider the distance between separated material things that is determined in terms of a concrete material reference frame, i.e. we are founding distant simultaneity on a previous notion that there are actually spatiality separated things whose distance can be measured in terms of an adopted spatial metric unit. Here is a version of how the argument is set forward: (1) according to Einstein "to determine the time at each point in space, we can imagine it populated with a very great number of clocks of *identical construction*" (Einstein 1910, 125);[13] (2) it is necessary to make sure that the clocks are synchronized, so that when for example determining the temporal behavior of a thing moving 'through' the 'net' of clocks, the reading of successive clocks really enables to determine the velocity between two successive points and make a meaningful reconstruction of the trajectory. According to Einstein "to get a complete physical definition of time ... we have to say in what manner all of the clocks have been set at the start" (Einstein 1910, 126); (3) as it is well-know, Einstein adopts a synchronization procedure (the one previously proposed by Poincaré; see, e.g., Brown 2005, 63) based on the propagation of light, which is not instantaneous like gravity in Newton's theory. According to Einstein we can regard different clocks is phase once the synchronization has been made (Einstein 1910, 127); in this way "the totality of the readings of all of these clocks in phase with one another is what we will call the *physical time*" (Einstein 1910, 127).[14]

4 Space in the theory of relativity

One of the aspects that differentiates the theory of relativity from Newtonian mechanics is in the prediction of a shorter length measurement of an extended material object in the direction of the relative motion of this object in relation to a reference frame from which this measurement is being made (see, e.g. Smith 1965, 63-74). What does this mean, and what are its implications regarding the notion of space?

The so-called length contraction is a kinematical effect resulting from any measurement procedure in which, e.g., the length of a rod is measured from a reference frame that is in relative motion in relation to the rod. From the perspective of a reference frame in relative motion the rod will appear to have a shorter length than what it actually has when measured in a reference frame in relative rest; If we bring the rod and the reference frame that are in relative motion to a state of relative rest we will find the same length as measured all the time by the reference frame that was always in relative rest. The actual length of the rod did not change only its measured or 'kinematic' length from the perspective of a moving reference frame (see, e.g., Einstein 1910, 128-9; Einstein 1911, 347-8; Reichenbach 1927, 195-7).

We know that in the Newtonian or Galilean space-time this situation does not occur; different reference frames (in the Galilean space-time) will give the same reading of the length of the rod (Logunov 1998, 15; Brown 2005, 38-9). This seems to be a big change regarding the notion of space; however a closer look makes possible a more circumspect reading of the situation. Centering our discussion in the Einsteinian space-time and Galilean space-time, in both there is no intrinsic notion of position as a 'property' of space. It is only by reference to a material reference frame that a definite metric can be attributed to what in the lack of a better word I have called spatiality. The *de facto* situation we face in the already relativistic Galilean space or the Einsteinian space is that it is only with a reference frame that we 'complete' our definition of space attributes; in this case that of a metrical distance between things. As we have seen, in Einstein's thinking one considers a spatial homogeneous vacuum, this sort of primeval spatiality, where the material reference frame (constituted by physical systems, like meter sticks and clocks) is immersed/part of; this same situation occurs with the Galilean space. As we have seen in the Galilean space one is left without a notion of absolute position; and following Einstein and dropping the ether one is also left without a putative substitute for the absolute space, in the sense that the ether might provide a privileged reference frame. The reference frame has in the Galilean space the same role as in the Einstein space, that of embodying the metrical aspects of spatial relations. To say it bluntly both the Galilean space, where one adopts the principle of relativity as applied to mechanics, or the Einstein's space, where one adopts the principle of relativity as applied to electrodynamics, share a non-metrical spatiality (that in the construction of the theories is logically previous-to-things), on which one completes the description of a space-scale by reference to material reference frames. Here we face an entanglement of the spatiality and material things in the completion of the so-called metrical 'properties' of space; Nevertheless it is possible to make a philosophical reconstruction of the situation so that one can for example defend a substantivalist view on space and time in the theory of relativity (see, e.g., Norton 2008); I will not go into these discussions here. I only want to stress that: (1a) logically, in the construction of the theory, the homogeneous vacuum comes before the reference frames in setting forwards the metrical 'properties' of space; (1b) conceptually, in the construction of the theory, one considers first an homogeneous vacuum where one situates physical systems like meter sticks and clocks; (2) operationally it is with the material reference frame that one can make measurements; (3) both the Galilean space and the Einsteinian space share this same framework for the construction of the metric on the primeval spatiality.[15]

---

13 Even if Einstein still speaks of a 'point in space' (Einstein 1910, 125), this must be read in the context of our previous discussion regarding Galilean spaces: a 'point in space' refers to a physical event located in relation to a material reference frame, i.e. to a practical situation in which a distance measure can be made (Einstein 1911, 340).

14 This spreading of time through space, or more exactly spreading of synchrony between clocks in different positions in a material reference frame has well-known implications regarding the notion of simultaneity of distant events. Due to the finite velocity of light, the synchronization procedure depends on the reference frame where it is made, i.e. it is relative to the choice of reference frame.

In a nutshell: There is a difference between the Galilean mechanical relativity and Einstein's relativity regarding measurements of length made in different reference frames in different states of relative motion. This does not have to be seen as a deep aspect of space; since the framing of the metric is dependence on physical things; one can choose to related it not to space but to, for example, the type of physical interactions that enable the measurability of lengths of physical systems in relative motion; in the case of the Galilean framework since we consider the case of an action-at-a-distance, there will be no difference in measurements made from different reference frames in relative motion; in the case of Einstein's framework we have a delayed interaction which will result in the kinematical contraction of length measurements made from different reference frames. What they have in common is an almost syncretic and vaguely conceptualized notion of spatiality, Einstein's homogeneous vacuum that we can see as present in both approaches. In my view it is in this shared primeval spatiality where the 'mystery' of space is; and we may loss sight of it when centering or argumentation and conceptual approaches on more elaborated and mathematically involved notions of metrical space (for example the four-dimensional manifold approach to the Galilean space-time or the Minkowski space-time) that incorporate different conceptual elements that we are not obliged to see strictly as properties/aspects of space.

---

15 While the view being defended here has a constructivist 'flavor', I am not trying to provide any sort of argumentation related to the so-called relationalist-substantivalist debate regarding the 'nature' of space. On this matter see, e.g., Earman (1989). Regarding a 'constructive' approach see, e.g., Brown (2005); Brown and Pooley (2006).

## 5 The so-called time dilatation and time in the theory of relativity

A well-known result of the theory of relativity is that different reference frames may give different readings of the duration for example of a recurrent physical process depending on their relative state of motion in relation to the physical process in question (see, e.g., Smith 1965, 49-60). This situation does not occur in the Newtonian or Galilean description. Does this imply a radically new conception of time in the theory of relativity?

Let us consider several clones spatially separated in a state of relative rest. How do we consider time with respect to these identical clones in the theory of relativity. We make use of a new concept related to time set forward by Minkowski, that of proper time (Torretti 1996, 96; Savitt 2011, 25). When one considers the spreading of time through space with resort to a reference frame, the typical approach, following Einstein, is to think in terms of a framework of identical clocks (see, e.g., Einstein 1911, 344-5); these clocks enable to measure the coordinate time in this reference frame, i.e. the time at a particular space location (defined relative to the particular adopted reference frame). Each clock embodies, so to say, a flow of time a 'moving' now; identical clocks will fell/have/embody the same temporal flowing; this is conceptualized in the theory of relativity by resort to the idea of proper time: each clock marks its own proper time. With Minkowski the central concept to think time in the theory of relativity goes from the coordinate time to the proper time which we can regard as logically, conceptually, and operationally more fundamental, and from which it is made the definition of the coordinate time. It is with this central concept that I will make and analysis of the conception of time in the theory of relativity.[16]

Each one of the spatially separated identical clones has the same proper time. When in relative rest near a clock of the reference frame of another clone, the proper time measured by the 'self-clock' (or 'internal' clock) of each clone will agree with the coordinate time measured by the (proper time of the) clock that is nearby. However when in relative motion the measurements of time intervals made by the proper time of a clone and the measurements of the corresponding time intervals made by the coordinate time of another clone do not coincide: this is the so-called time dilatation effect. Like in the case of the contraction of length measurement this is a kinematical effect. What does this implies? Let us consider the measurement of the proper time of each clone made from the other clone in relative motion. All of them 'live' the same temporal flow; however they attribute to each other different temporal flows, i.e. when in relative motion each clone will attribute to the others a smaller temporal flow, i.e. they will look younger; time seems to go slower to them.[17] However if now all the clones come to relative rest they will see that they are of the same age, no one is older than the other and again the coordinate time of each clone will show that the other clone has the same aging.[18]

---

16 Minkowski original definition of the proper time of a 'substantial point' was made by resort to an already defined reference frame with its coordinate time (Minkowski 1908). In fact this is the current procedure in which the proper time of a material system is determined in terms of the coordinate time of an inertial reference frame (Bohm 1965, 161-4). The view presented here is that when one is talking about the spreading of time through space (i.e. the spreading of distant synchronicity), one is already implicitly using each clock of the reference frame as a 'bearer' of a proper time and arriving at a coordinate time by synchronizing these proper times.

17 This can be easily understood by taking into account the fact that while each clone measures its own proper time with just one clock (i.e. it measures a proper time interval), when measuring the other clone's proper time, each clone needs two clocks (i.e. it measures an improper time interval). Both clones when measuring their own proper time arrive at the same result and both of then attribute to the other the same slower time flow (see, e.g. Smith 1965, 49-60). As Einstein called the attention to "every event in a physical system slows down if the system is set into translational motion. But this slowing occurs *only from the standpoint* of a non-comoving coordinate system (observer)" (Einstein 1915, 257 [my emphasis]). Like in the case of the so-called length contraction Einstein refers to the so-called time dilatation as a "purely kinematical consequence of the theory" (Einstein 1911, 348).

18 Regarding this last point one must take into account an important assumption implicit in the construction of the theory. According to Einstein, "we will always implicitly assume that the fact of a measuring rod or a clock being set in motion or brought to rest does not change the length of the rod or the rate of the clock" (Einstein 1910, 130). This is called by Brown the boostability assumption (Brown 2005, 30). Also it is important to notice that we are not considering in this case any accelerated motion by any one of the clones, only states of relative motion. As it is well-known, if there is an accelerated motion by one of the clones, this clone will actually be younger (see, e.g. Smith

Like in the case of the length contraction one does not have to be seen in the time dilatation any deep aspect of a newer conception of time; one can choose to related it not to time but to the measurement of a time interval made by an observer in relative motion for which the finite velocity of light is a key element (see, e.g., Smith 1965, 50-7). In the case of an infinite velocity of light (i.e. action-at-a-distance interactions) there would be no difference regarding time interval measurements made by different observers in relative motion (in operational terms, if not conceptually, we would be in the same situation as in the Newtonian or Galilean space-time). However there is a new actor in stage, the proper time, where there might be a crucial change in the conception of time in the theory of relativity; also, one knows that there seems to be more changes regarding time in the theory of relativity that arise due to the so-called relatively of simultaneity.

6 The relativity of simultaneity: a first look

As we have seen, according to Einstein, the relativity of simultaneity is the more challenging proposition of the theory of relativity in what regards our previous conception of time. Strictly speaking the relatively of simultaneity is just one of several interrelated aspects of the theory of relativity, it comes hand in hand with the so-called length contraction and time dilatation; they are all 'codified' in the Lorentz transformation (see, e.g., Einstein 1915, 254-7; Stephenson and Kilmister 1958, 37-42). However it is with the relativity of simultaneity that previous notions related to time seem to be on the spot. In fact there are notions related to the conceptualization of time like now, becoming, flow, tense (i.e. the differentiation between past, present and future) that seem to be at odds with the theory of relativity (see, e.g., Savitt 2011).

The usual way to stress this situation is through the conceptual construct called event. An event consists in a related distance or length measurement and a time interval or duration measurement abstracted as a four-dimensional point corresponding to a precise location and moment in time. Usual examples of events are the sending of a light signal at a precise moment from a particular point (in relation to a reference frames where this location is defined) or the striking of a thunder at a particular moment and at a particular location (according to a reference frame where space and time coordinates are defined). If we now consider two events that occur at the same time in a particular reference frame, i.e. events, occurring at a distance $\Delta L$ and for which $\Delta t = 0$; from the perspective of another reference frame the events will occur at different moments corresponding to a non-zero time interval between them $\Delta t´ \neq 0$, and also a different distance $\Delta L´ \neq \Delta L$ (see, e.g., Stephenson and Kilmister 1958, 37). Another away of arriving to the same result is by looking into the coordinate time of each reference frame. In each reference frame the clocks are synchronized, i.e. their identical proper time provide a spatially located (coordinate) time; now, from the perspective of a reference frame in relative motion (that has its own clocks synchronized) the clocks of the other reference frame are seen as desynchronized (see, e.g., Einstein 1905, 145; Bergmann 1942, 38). In this way, for one observer two thunders may be said to strike at the same time, now. However another observer with an identical proper time will nevertheless consider that, for example, one thunder strikes some time ago in an earlier now and the other is striking in a later now.[19] It seems that if we try to use the thunder strikes (i.e. the events) to provide a notion of now

---

1965, 93-102).

19 We can also consider this issue from the perspective of Minkowski diagrams (see, e.g., Bohm 1965, 162; Schutz 1985, 16). If one thinks in terms of a particular physical system O one can 'replace' its successive 'nows' marked by its proper time by a worldline (i.e. a continuous set of events that in general represent in terms of the coordinate time the successive spatial location of O in any reference frame). This is the simplest case, that of O's 'own' reference frame in which we represent its worldline as a vertical line locate at the spatial origin of the reference frame (i.e. the worldline is the time axis). The event representing the here and now of say a human observer is represented as the apex of two so-called light cones; all the events lying in the future light cone are later than the event representing the here and now of an observer and they are so in any reference frame; in an equivalent way all the events lying in the past light cone are earlier than the event representing the here and now, and they are so independently of the reference frame. The cumbersome aspects of the relatively of simultaneity appear in relation to the events that lay we are in trouble. This situation seems to be clearly at odds with the Newtonian or Galilean notion of time in which even in the case of a Galilean space-time it seems that every event that happens now for one observer also happens now for another observer in relative motion with the first. All observers share the same now and all simultaneous events also. As Einstein mentioned, the relatively of simultaneity seems to indicate the need for a throughout revision of basic aspects related to the concept of time.

7 On Dieks views regarding time in the theory of relatively

According to Dieks the theory of relatively teaches us that it is not necessary to rely on an idea of a succession of cosmic nows; in his view, " if we want to make sense of becoming we should attempt to interpret it as something purely local" (Dieks, 2006, 157); one must consider the successive happening of physically related events from the perspective of "their own spacetime locations" (Dieks 2006, 157). This points to the centrality of the concept of proper time in the theory of relativity.

As it is well-known, only events on the past light cone can affect us, and only events in the future light cone can be affected by us (see, e.g., Callahan 2000, 76-7); all the events that are spacelike separated from us, since there is no action-at-a-distance, cannot have a causal influence on us (i.e. there cannot be any physical interaction with them). Accordingly they "have no influence on the content of our observations" (Dieks 2006, 158), in particular the ones that helps us in the construction of a space and time conception. Dieks conclusion is quite interesting:

we do not need a succession of a definite set of global simultaneity hyperplanes in order to accomodate our experience ... completely different choices of such hyperplanes lead to the same local experiences ... we do not have to bother about global simultaneity at all. If we decided to scrap the term 'simultaneity' from our theoretical vocabulary, no problem would arise for doing justice to our observations. (Dieks 2006, 160)

The importance of the relatively of simultaneity would not then be directly on the time ordering of distant events but in pointing to the crucial aspect that the temporal experience is local.

Dieks proposes to reformulate the idea of flow of time based on the concept of proper time. According to Dieks 'only time *along worldlines* (proper time) has an immediate and absolute significance as an ordering parameter of physical processes' (Dieks, 1988, 456). However since there is not in the theory of relativity any preferred worldline (or associated reference frame, as is the case of an ether based reference frame in Lorentz's electron theory), there is no way to single out a particular worldline and its private now. Accordingly, 'it is not appropriate to define one universal 'now'; instead, we have to assign a now-point to every single worldline' (Dieks, 1988, 458). Dieks view is that, contrary to earlier views on time, the relativistic framework leads to a generalization of the universal flow of time to what Dieks refers to as flow of time per worldline. However, this view might lead to a consistency problem. According to Dieks an arbitrary assignment of now-points to the worldlines will not do, however, for the following reason. The idea of a flow of time combined with the ontological definiteness of present and past requires that everything that is in the past lightcone of an event that is ontologically defined is also ontologically define. This leads to the demand that no now-point should lie in the interior of the conjunction of the past lightcones of the other now-points. (Dieks 1988, 458)

The question is how do we obtain a consistent assignment of now-points? We consider it as a sort of initial condition put by hand?

8 'Now' as the spatial co-existence of physical systems or processes

---

outside the light cones (where there is a so-called spacelike separation), for these, like our example of the two thunders, depending on the reference frame they might be felt as happening now by one observer our at different nows by another (see, e.g., Bohm 1965, 150-4).

The theory of relativity enables us to compare space and time determinations (i.e. measurements) made, in particular, by different inertial observers;[20] that is, it provides a way (the Lorentz transformation) to compare the coordinate systems of different observers (i.e. a way to related each other reference frames). As already mentioned, a reference frame associated with a particular physical system can be idealized as a sort of space filling grid of identical measuring rods and clocks. All these identical material measuring rods and clocks (all of them physical systems) co-exist spatially with our appointed observer (a particular physical system with a particular proper time that we consider to be at the origin of the reference frame), i.e. they inhabit the same homogeneous spatiality; there is no question of thinking that they might be spatially located in relation one to another while being in different nows. A clock cannot be, in the reference frame, one measuring-rod away from the observer, and not being co-present with the observer. If this clock was in the past or future of the observer it would not be actually there in the reference frame. The clocks of the reference frame are present, or better co-present with the observer because we consider all of them to co-exist in space. Here, space must be understood in the previously mentioned sense of homogeneous vacuum; a sort of pure spatiality previous in the conceptual construction (both in Galilean and Einsteinian space-time) to the reference frame that enable to frame a metric in this yet not metrical spatiality.[21]

In the theory of relativity all physical systems or processes are represented in the reference frame as timelike or lightlike wordlines. The timelike wordline corresponds to physical systems in part made of matter (even if they are complex and involve, e.g., electromagnetic interactions). The lightlike worldline is the worldline of 'light' (i.e. electromagnetic radiation, since in the theory of relativity there is no treatment of gravity or other 'fundamental interactions'). The clocks in the reference frame must be visible between them, i.e. they can interchange light signals and in this way check if they have the same proper time. In other words each timelike worldline of the reference frame must be connectable with other timelike worldline by a lightlike worldline. The same can be said regarding the worldline of any other physical system as seen from the reference frame under consideration: they must be visible in the reference frame. This means that *implicit in the theory of relativity is the idea that an inertial reference frame fills all of the homogeneous vacuum*, or at least that the only physically meaningful notion of space is the one derived by spatial distances that are measurable within a reference frame.[22]

Since there is visibility between the clocks of the reference frame one can crosscheck each clocks's proper time from the perspective of another clock proper time. This can be done by sending a light pulse in the end of each clock's cycle to the other clocks and measuring in these clocks the time interval for the arrival of successive light pulses.

---

20  The theory of relativity enables besides the description of inertial reference frames and the relation between coordinate systems defined within them, also the description of non-inertial reference frames (see, e.g., Misner, Thorne, and Wheeler 1973, 163-74; Logunov 1998, 135-46; Friedman 1983, 132-5).

21  Having defined a reference frame we can think of each identical clock of the grid in terms of a temporal worldline; in this case they will be represented in a Minkowski diagram as vertical lines (taking the observer's worldline to be positioned at what we take to be the origin). It is implicit in the construction of the reference frame that the clocks have an identical proper time, and that we can check it out.

22  We can see this idea of visibility present in physics even previous to the theory of relativity. For example Kepler objected to an infinite space on account that the stars that are visible by us cannot be at an infinite distance, and that eventual stars not visible are not empirically accessible (Koyré, 1957, 76-87). Within the theory of relativity, Reichenbach comments that to be able to compare two time series at different points in space, "we must first make the assumption that there always are connecting signals between any two points in space" (Reichenbach 1927, 143). That is, physical systems located (within a reference frame) at different points must be able to exchange signals. Here it is implicit the above mentioned idea that the reference frame fills all space, or at least that we only have a meaningful notion of metrical space from the perspective of a reference frame: we only talk of a point in space by first defining a reference frame where this point is meaningful (we do not have anymore an absolute space with its own self-positions), and these 'points' must be visible by construction. Also Whittaker mentions that to be able to talk about the spatial distance between two physical systems it is necessary that they are connectable by a lightlike worldline (Whittaker 1953, 186).

Let us now consider another physical system (an observer) immersed within the reference frame, i.e. a physical system co-existent with the measuring rods and clocks of the reference frame. We want to address the question of its worldline now-point. Let us consider the observer to be motionless within the reference frame. We can consider that this is its own reference frame. In this case there is no questioning regarding the sharing of the same now-point between the observer and the rest of worldlines of the reference frame. However let us now consider that the observer is in relative motion within the reference frame. We can only trace this system's trajectory in the reference frame if we can interact with it; in Newtonian or Galilean space-time one considers a direct physical interaction at a distance; in the theory of relativity one considers a delayed interaction *at a distance*. Nowhere are we using a concept of an interaction into the past or into the future; it makes no sense: the interactions are spatial. Only if the physical system is there spatially within the reference frame, visible in it, can one measure, e.g., its motion using measuring rods and clocks. To say in another word the physical system has to be immersed within the reference frame, i.e. *the moving physical system is co-existent with the measuring rods and clocks of the reference frame: the physical system is present to the other physical systems of the reference frame*.

This same view can and must be extended to the reference frame attached to the physical system. It must be present within the other reference frame.[23] How could we compare measurements of durations and distances if it was not the case? Let us look at this in detail. I will follow Einstein here. He asks us to consider two equivalent reference frames that have the same measuring rods (i.e. the same unit of length) and clocks that run in synchrony when the frames are at relative rest (Einstein 1910, 130); I will consider the clocks to be identical physical systems with an internal clock that measures their identical proper time. Both reference frames are co-present, they co-exist in space. Now one considers that the two reference frames are set into relative motion.[24] A motion is not slipping into the past or the future. It is always happening in a common now. Each reference frame is moving as a block in the perspective of the other reference frame as time goes by.

Let us consider two clones A and B in the origin of each reference frame, and that the path of clone A passes just next to the clone B. At this moment one can set the time of each clone to zero,[25] i.e. one sets the time in each reference frame to zero.[26] Let us suppose that at the moment the two origins of the reference frames (i.e. the two clones) coincide a light signal is send. The velocity of light is the same in both reference frames;[27] this means that the spatial positions x, y, z in frame A

---

23 Logically this is immediate: (1) the physical system A and its associated reference frame A (with measuring rods and clocks) are present to each other; (2) the physical system B is, e.g., momentarily passing by the origin of the reference frame A, next to the physical system A, i.e. it is present to A; (3) since B is present to A, its reference frame is also present to A and to A's reference frame.

24 As already mentioned, Einstein notes that "we will always implicitly assume that the fact of a measuring rod or a clock being set in motion or brought to rest does not change the length of the rod or the rate of the clock" (Einstein 1910, 130; see also Brown 2005, 30).

25 Since one considers that the clocks have been previously synchronized in each reference frame this sets the time for all the clocks of both reference frames to zero; better, what one does is as follows: let, e.g., A' clock be reading $t_A$ when it encounters B as all other clocks in the reference frame of A that have been synchronized with A's clock; let B's clock be reading $t_B$ as all other clocks in B's reference frame. One does not set A's clock to zero and synchronized again, and equivalently in B's frame; one simply subtracts $t_A$ to the subsequent readings made by clocks in A and subtracts $t_B$ to the subsequent reading made by clocks in B. It is important to notice that we take the clocks to be synchronized previous to the derivation of the transformation equations. I will return to this point below.

26 It is usually considered that the origins of the coordinates systems coincide for $t = t´ = 0$. More generally one associates the coordinates $x_0$ and $t_0$ to $x´ = t´ = 0$, which are inbuilt in the Lorentz transformation as additive constants (see, e.g., Stephensen and Kilmister 1958, 11). What this means is that when we set to zero the clock located at what we take to be the origin of, e.g., the reference frame A (usually associate with a special physical system, we call the observer A) in the perspective of the reference frame B the observer is momentarily located just next to a specific clock of B marking a time $t_0´$ and located $x_0´$ measuring-rods away from the origin of reference frame B.

27 This is the so-called principle of the constancy of the velocity of light, which is an abbreviation of saying that, e.g., "a ray of light in vacuum always propagates with the same velocity c, which velocity is independent of the motion and x´, y´, z´ in frame B that are just reached by light at the time t (measured by a clock located at x, y, z in frame A) and time t´ (measured by a clock located at x´, y´, z´ in frame B) are giving by $x^2 + y^2 + z^2 = c^2 t^2$ and $x'^2 + y'^2 + z'^2 = c^2 t'^2$. The transformation equations between the two coordinate systems (made possible by the two reference frames) must enable to transform each of the previous equation into the other.

We are considering a homogeneous spatial vacuum substantiated in the construction of the theory in that there is no privileged position neither any distinction of direction when defining a reference frame;[28] also in the theory of relativity the rate of a clock (its proper time) is taken to be uniform.[29] These pre-conditions to the construction of the theory of relativity imply that the transformation equations must be linear.[30] By requiring that the inverse transformation from B to A should give us again the coordinate system of A, one arrives at the so-called Lorentz transformation.[31]

The simple aspect I want to call attention to is that *to arrive at the Lorentz transformation one must consider two inertial reference frames co-existing in space*. For example the clone B located at the origin of reference frame B is always present to the reference frame A in which B's trajectory can be measured using A's measuring rods and clocks.

A critic might say at this point that even if one accepts the idea of spatial co-presence (i.e. co-existence) this is not the same as temporal co-presence (i.e. the possible identity of the now-points of different worldlines), since what is at issue is the relation between the 'internal' proper time of one system with another. In fact one can agree that a spatial co-presence is not disputable since it is a presupposition for the possibility that two physical systems might have each one a reference frame and may use these reference frames to reconstruct the worldline of the other system. So a critic might agree that spatial co-presence is a necessary presupposition adopted in the theory but not that it implies a temporal co-presence. In what follow I will argue that in fact a notion of temporal co-presence is possible in the theory of relativity (i.e. that physical systems in fact share

---

of the body that emits the ray" (Einstein 1910, 124). In presentations of the theory of relativity this principle and the principle of relativity are usually given as two independent principles from which the theory is built (see, e.g., Smith 1965, 41; Pauli 1958, 5). However the relation between the two principles is more intricate. As Einstein mentions the principle of the constancy of the velocity of light must be compatible with the principle of relativity (Einstein 1910, 124). It is not an independent principle, but at the same time it is not just one more consequence of the principle of relativity. According to Paty, "le second – le principe de constance de c – est soumis au premier, mais il fournit à celui-ci les nouvelles définitions conceptuelles qui permettent de l'exprimer dans son universalité" (Paty 1993, 143). In Tolman's view, "this postulate can be looked at as the result of combining the principle familiar to the ether theory of light, that the velocity of light is independent of the velocity of its source, with the idea resident in the first postulate which makes it impossible to assign any significance to the absolute velocity of the source but permit us to speak of the relative velocity of the source and observer" (Tolman 1934, 15).

28 I will not go into the discussion of the possible conventionality (and its consequences) of the adoption of a Cartesian coordinate system and the so-called one-way velocity of light (see, e.g., Friedman 1983, 294-320, Janis 2010; Brown 2005, 19-22 and 95-98; Darrigol 2007; 536).

29 This is usually taken to be a convention (see, e.g., Poincaré 1898, 2-3; Reichenbach 1927, 114). Again I will not discuss the possible implication of the (eventual) conventionality of the uniformity of time.

30 Friedman criticizes this type of derivation of the Lorentz transformation on account of an, in his view, imprecision in the meaning of homogeny of space and time (Friedman 1983, 140-1). Even if Einstein did not give much details regarding the relation of homogeneity and the linearity of the transformation equations he gave same remarks that can help us understand the issue (see Einstein 1910, 137; Einstein 1912-1914, 31); also, it is possible to make mathematically more precise what one means by the homogeneity of space and time (see, e.g., Dutta, Mukherjee, and Sen 1970; Brown 2005, 26-8). In particular in what refers to the homogeneity of time, since we think in terms of an underlying homogeneous vacuum there cannot be any difference, resulting from the transformation, in the proper time of a clock depending on where it is locate within a reference frame A, neither there should appear a difference, as result of the transformation, between the proper times of identical clocks that are at rest and located at a distance in the reference frame A; also the rate of the clock as measured in a reference frame B cannot be changing in time depending on the clock's time in the reference frame A, i.e. *time is taken to be uniform, or in other words homogeneous*.

31 By considering the simplified case where time is set to zero when the origin of the reference frames coincides and that the relative motion is taken to be in the x-direction, the Lorentz transformation is given by: $t' = \beta(t - v/c^2 x)$; $x' = \beta (x - vt)$; $y' = y$; $z' = z$, where $\beta = 1/\sqrt{1 - v^2/c^2}$, c is the velocity of light, and v the relative velocity between the two reference frames.

the same now-point, contrary to Dieck's view), and that this notion boils down to the notion of spatial co-presence or co-existence. That is I will defend that *what we call now or present is a spatial aspect, not a temporal one*.

Let us return to the relation between the two reference frames; how can we cross-check the now-point of the two co-existing observers located at the origin of the reference frame?[32] I will consider that each clone goes through a specific cyclic process (or has an internal clock or an adjacent atomic clock) that enables to measure the flow of its own proper time. To enable clone B to measure the time flow in A, one considers that A is sending a light pulse to B at the beginning of each of its internal cycle. Let us consider the case in which B is located at a specific position in A's reference frame, i.e. A and B are in relative rest. Since B and A are identical physical systems they go through identical physical processes, i.e. they measure time with identical 'internal' clocks going through an identical flow of time. Supposing that A and B's clocks are synchronized to the same phase (e.g. $t = t' = 0$), B will receive A's light $d/c$ seconds after it is send (where d is the distance between A and B as measured in A or B reference frame, using measuring rods specified as the unit of lenght, and one considers/defines that one internal cycle of the clones takes 1 second; c is the speed of light measured using this system of units); if the second pulse is send $\Delta t$ seconds after the first in A's clock it arrives at B at $d/c + \Delta t$ seconds in B's clock. In this way, both A and B attribute a time interval of $\Delta t$ to the cycle A goes through, i.e. they measure the same time flow. At this point there is no reason/justification to make a principled distinction between (temporal) now from the spatial co-existence of physical systems or processes.

The situation of a relative motion between the two clones is more involving. Already within a Newtonian or Galilean space and time one knows that, due to the Doppler effect, B (A) will attribute to A (B) a different proper time than it actually has: if clone A emits pulses $\Delta t$ seconds apart, these are received by clone B (in relative motion away from A in the x-direction at speed v) $\Delta t' = \Delta t (1 + v/c)$ seconds apart (see, e.g., Smith 1965, 25). If one did not took into account the Doppler effect one might be lead into a confusion regarding the notion of now: for B a second of A seems to take $1 + v/c$ seconds to pass. This might give the impression that A and B, even if co-existing spatially and interacting one with the other (e.g., by exchanging light pulses), might have their now-points going out of 'synchrony' as if A 'arrives' first into the future.[33] That does not occur. In fact the Doppler effect is a kinematical effect related to a measurement of a wave in which there is a relative motion between the wave source and the observer. By taking into account the Doppler shift of the observed frequency of the successive pulses one knows that B has to compensate (i.e. calculate away) the shift so that it arrives at the actual time interval of emission of the pulses by A: *if A emits pulses one second apart of its proper time, and B measures the time interval of the successive pulses using its proper time and takes into account the Doppler shift, B will attribute to 1 second of A's proper time one second of B's proper time*. In this way time is flowing equally for both clones.

In the case of the theory of relativity one has to consider besides the Doppler effect the time dilation effect, again a kinematical effect as we have seen. In the relativistic case, clone B measures, with its proper time, a time interval $\Delta t_B = \Delta t'_A (1 + v/c)$. The crucial difference to the non-relativistic case is that in the previous expression $\Delta t'_A$ (the time interval between emission of successive pulses by A) is being measured in B's reference frame, i.e. it is an improper time interval; in terms of a proper time interval measured by A's clock (i.e. in terms of A's proper time) $\Delta t'_A = \Delta t_A / \sqrt{(1 - v^2/c^2)}$. In this way if A emits pulses 1 second apart (of its proper time), B will receive them (1 + v/c)/ sqrt (1 − $v^2/c^2$) seconds apart (of its proper time). If B does not take into account the Doppler and time dilatation effects it will consider that time flows differently to clone A, i.e. that they are aging differently.[34] If now the two clones come to relative rest they will again measure the same proper time, since there are no kinematical effects anymore affecting the measuring procedure.

---

32 I will consider here the now-points of the worldlines of clone A and clone B; but the same can be done in relation to any worldline of physical systems belonging to each reference frame. Since all the clocks and measuring rods of A, at rest in relation to A, are temporally co-present with A and the same goes for B, one can expect that if clone A and B are in the same now also all the other physical systems or processes occurring within the two reference frames will also share the same now. To show this clearly I will have to address the so-called relativity of simultaneity. This is done below.

33 I will show below that a situation in which actually there is one clone that has a smaller flow of time, the case of an accelerated physical system, does not have to lead to this type of conclusion.

We are considering an initial situation where the origin of the two reference frames coincided.[35] The two clones synchronize their clocks at this moment, their common now. When they come to relative rest each clone will be located in the other reference frame next to a clock, which can check that the proper time is the same (or they can interchange light pulses to confirm this). They are now with this clock (i.e. they are now with the other clone and its reference frame). Could we say that during the relative motion each clone has a different now-point even if at the beginning and at the end of the relative motion they have the same now-point? Or asking in another way: when in relative motion is there a need for a notion of now that is not encapsulated in the notion of spatial co-existence?[36]

Let us consider that clone A sends pulses 1 second apart (measured in its proper time) to clone B that is in a relative motion away from A in the x-direction in A's perspective. Let us say that A and B set their clocks to zero 6 seconds ago; this is now for A. We will consider that at the moment that B receives a light pulse from A it send back another light pulse (or that this is done by the clock of A's reference frame locate momentarily next to B). Since light takes time to arrive at A, A cannot know what is happening to B in A's now, only what is in A's past light cone is known by A.[37] We can only make an educated guess based on our knowledge of the behavior of the physical system B.[38] Let us consider that in what for A is now, A and B actually come to relative rest.[39] A will know that a few seconds later (of its proper time). If clone B looks at a clock from A' reference frame nearby, it will see that the clock is marking 6 seconds, since we take all clocks from A's reference frame to have the same proper time and to have been synchronized so that they have the same 'initial phase'.[40] However by measuring the pulses arriving from A, B will attribute to A, as we have seen, a different passing of time. It is only if B does not have access to the clocks in A's reference frame that it can confuse its kinematical measurement of an improper time interval with A's actual proper time (as measured by a distant clock in A's reference frame that has been synchronized with A's clock).

Even if B does not have direct access to the time readings in A's reference frame, it is presupposed that B can make measurements of A in its reference frame, in particular B can know A's trajectory. In this way B knows the direction and velocity of clone A in its reference frame. This means that B can calculate away the kinematical effects (i.e. the Doppler effect and the time dilatation effect) of its measurement, using its proper time, of the time that A went by until the moment B and A came to relative rest. By doing this calculation, B determines that A passed by 6 seconds of B proper time, i.e. the same time interval measured by A with its proper time. We can check this easily by considering that B was also sending light pulses to A one seconds apart of its proper time. According to the principle of relativity we face a completely symmetrical situation if we consider the motion of A from B's reference frame.[41] According to B it is also at the $6^{th}$ second of its proper time that A comes to relative rest. Again we arrive at the result that when the relative motion is over each clone went by the same amount of time (i.e. they have the same age), and that their clocks are still synchronized.[42] What happened regarding each clone now-point when they were in relative motion? If both of them went through 6 second of their proper time and one presupposes the uniformity of time (or in Einstein's words the homogeneity of time), it is immediate to consider that when clone A went through 1 second (and send to B a second light pulse) the same happened to clone B, i.e. B went also through one second (of its proper time, that is identical to A's proper time). It is only if the clones do not take into account the kinematical measuring effects that they might wrongly attribute during relative motion a different time passing to each other and might double-wrongly regard this as implying that they had different now-points. In this way *when considering relative motion there no need for a notion of now that is not encapsulated in the notion of spatial co-existence.*

---

34 If this was the case this would imply a breakdown of the principle of relativity, since physical systems in relative motion would behave differently, in this case in what regards the passing of proper time.

35 As we have seen we do not need to consider this simplifying assumption to establish the Lorentz transformation. See footnote 27.

36 It is undisputed that while in relative motion each clone is moving within the other's reference frame, i.e. it is co-existent with the other clone and its reference frame.

37 At the moment B receives a light pulse from A it is located near a particular clock from A's reference frame marking a determined time. We can consider that this clock registers this particular moment and that this is done by all the clocks B is passing by in A's reference frame, i.e. one is registering B's trajectory in A's reference frame. If the clocks in B's trajectory send back the information to A, this cannot be done at a velocity superior to that of light.

38 When analyzing this type of problem with the help of Minkowski diagrams one has the tendency to use the diagrams as if one is outside space-time not representing any now-point, as if there was no meaningful notion of past, present and future. As Bohm called the attention to, a space-time diagram is always a reconstruction of past events, i.e. related distance and duration measurements made using an actual reference frame (Bohm 1965, 173-178).

39 Here by the boostability assumption one considers that A and B's proper time are not affected by coming to relative rest. See footnote 18.

40 In deducing the Lorentz transformation it is made an explicit use of the possibility of synchronizing two clocks moving pass each other: the usual setting of t = t´= 0 at the origin; This implies the possibility of an observer actually 'seeing' the other observer's clock reading. In fact one can imagine that a photography is taken at the moment the clone B is passing by a particular clock from A and that in this way we have access to B's time and also to A's corresponding time (see, e.g., Smith 1965, 54).

That we would by facing a double misconception regarding the notion of now can be seen by considering the so-called twin paradox (see, e.g., Smith 1965, 93-102). Here we face an asymmetrical situation in which one of the clones, lets say B, is accelerated. Let us consider B to have a uniform circular motion starting and finishing at where A is located. Again we take B to be an identical physical system to A (i.e. a 'clone' of A), and both to have their clocks synchronized before B starts its accelerated motion. It turn out that, at the end of the accelerated motion when B is back side by side with A in relative rest (or passing by A), if A and B compared their time readings, seeing, e.g., the positions of the hands of each other clocks they are not the same. This is completely different from the previous case of a relative motion between the clones in which at the end of the motion the 'internal clocks' of each clone give the same time reading (6 seconds in our example). In the case of an accelerated motion, clone B goes by less time than clone A: if clone A goes by T seconds of its proper time, clone B goes by T sqrt $(1 - v^2/c^2)$ of its proper time.[43] Here we do not have any kinematical effects regarding the measurement of time intervals. Does this imply anything regarding the now-points of each clone? Since there is actually a different flow of (proper) time for each clone one might be lead to think that the relation between the now of clone A and the now of clone B, at least, during motion cannot be described as the sharing of a co-present simply due to the fact of being spatially co-existent.

---

[41] We see this immediately in terms of Minkowski diagrams. The Minkowski diagram representing in A's reference frame the sending of light pulses to B is symmetrical to the Minkowski diagram representing in B's reference frame the sending of light pulses to A (see, e.g. Bohm 1965, 134-7).

[42] Since they were synchronized at the beginning of the relative motion, and they are identical physical systems that having the same proper time (i.e. identical internal clocks) went through the same amount of time, if we take into account the bostability assumption one does not consider that after a relative motion it is necessary to synchronize again the clocks of the two reference frames, i.e. clone A and B share the same simultaneity plane.

[43] To arrive at this result, that can be confirmed experimentally (see, e.g., Zhang 1997, 175-200), one must consider a sequence of co-moving inertial reference frames where the accelerated clone is momentarily at rest; in this way one can see the accelerated motion of clone B as a continuous passing from one instantaneous co-moving reference frame to another (Logunov 1998, 149). The trajectory of clone B in A's reference frame is a temporal-like worldline. In A's reference frame the origin of a co-moving reference frame will move d**r** = **v**dt during the infinitesimal time interval dt where we take the accelerated clone to be 'almost' at rest in the origin of the co-moving reference frame. In the co-moving reference frame, clone B is momentarily at rest during the time interval dt´ *as measured by clone B's proper time*, i.e. in the co-moving reference frame one reads the time interval of the accelerated clock (i.e. of clone B) that is at rest in this frame. Thus, due to the invariance of the interval $ds^2 = c^2dt^2 - d\mathbf{r}^2$, we have $dt´^2 = dt^2 - d\mathbf{r}^2/c^2$, i.e. dt´= dt sqrt $(1 - v^2/c^2)$ (see, e.g., Logunov 1998, 41; Bohm 1965, 163-4). In this way, *by considering only a co-moving reference frame and the relativistic invariance of the interval, one determines the behavior of the accelerate clock* (i.e. the 'internal clock' of clone B). By considering an infinite sequence of co-moving reference frames one arrives at the result: Δt´ = ∫ sqrt $(1 - v(t)^2/c^2)$dt. It is usually considered that to arrive at this result one has to rely on the so-called 'clock hypothesis' that entails that the rate of a clock does not depend on its acceleration (see, e.g., Bohm 1965, 163; Brown 2005, 94-5). According to the view defended here this 'hypothesis' does not seem to be necessary (for a similar view see Arthur 2010).

It is clear that just before and just after the accelerate circular motion of clone B, it shares the same now as clone A: they are spatially co-existent next to each other, and their internal clocks (i.e. their proper time) flow at an equal pace. Also it is clear that, during motion, clone B is always spatially co-existent with A, since we can trace in A's reference frame B motion within it. B 'embodied' time is flowing slower than A's, but B is always (i.e. at every moment of A or B's proper time) present in A's reference frame. Let us consider that when the clone B passes just next to a clock in the grid constituting A's reference frame, the clock not only registers the time but also registers the time being measured by B's internal clocks (e.g. it takes a snapshot of the hands of B clock). This can be done since B is visible in A's reference frame; we can not only recover B trajectory but also B's 'internal' clock phase while moving within A's reference frame. Clone B does not go into the past or future of A' reference frame, it is always there, co-present with A and its measuring rods and clocks. Again it does not seem necessary to take into account more than the spatial co-existence to have a notion of present.

This implies that the notion of present does not need a sharing of the same or related flow of time, i.e. it is not necessary that different physical systems share the same uniformity of time. What is necessary, to measure the non-uniformity of the proper time of accelerated systems, is that the accelerated clocks exist within a background of inertial physical systems that have a 'stable' proper time, i.e. that there is a constant rate between each other proper time that can be checked experimentally. In this way the uniformity of time is 'substantiated' in the stability of the different proper times of physical systems 'belonging' to inertial reference frames.

As we have seen, according to Einstein and others (see, e.g. Einstein 1914, 4; Savitt 2011, 23-5), the so-called relativity of simultaneity is the must challenging aspect of the theory of relativity regarding the previous notion of time. Let us look again at the relativity of simultaneity from the perspective of the proposed view of now as (just) the spatial co-existence of physical systems and processes. The different clocks located in the 'grid' of a reference frame are measured as having a slower pace from the perspective of a reference frame where the grid is in motion. *All of them, when in relative motion, are measured as having the same (slower) rate*. The so-called time dilatation is associated with all the clocks of the reference frame in relative motion with another from which the duration measurements are being made (and the same situation occurs in relation to the so-called length contraction). It does not seem that the time dilatation has a direct bearing on the question of the relativity of simultaneity since all the clocks face the same apparent change from the perspective of a reference frame in relative motion with the grid where we take the clocks to be located at.[44] In the derivation of the Lorentz transformation it was implicit that in both inertial reference frames the clocks had been synchronized. The fundamental aspect of the distant clock synchronization is that it is a procedure dependent on the reference frame where it is made. As Einstein called the attention to, two clocks that are synchronized in one reference frame (i.e. two clocks with identical proper time that are set on phase), are seen as being in a different phase by a moving observer (Einstein 1905, 145). In other words the a procedure that in a reference frame where the clocks are at rest is seen as a synchronization of the clocks phase (e.g. putting both clocks hands in the mark 12h) is seen from a reference frame in relative motion as a desynchronization procedure. This can be seen by analyzing the Lorentz transformation. In it we have 'codified' the so-called time dilatation, the length contraction, and the relativity of simultaneity. Let us consider a clock located at the origin of a reference frame. As we have already seen, from the perspective of a reference frame in relative motion, a time interval measured by this clock will seem longer, i.e. the clocks seems to be running slower. This result follows immediately from the Lorentz transformation $t´ = (v/c^2 x + t)/\sqrt{1 - v^2/c^2}$. We have then $t_2´ - t_1´ = (t_2 - t_1)/\sqrt{1 - v^2/c^2}$. What about the phase of the clock when seen by the reference frame in relative motion? Let us recall that we have considered that the clocks at the origin of each reference frame were actually put on the same phase (i.e. synchronized) when the origins of the reference frames coincided, i.e. we made t´= t = 0. For this clock we have x = 0; this means that in this case t´ = t/sqrt(1 – $v^2/c^2$). That is, we only have a time dilatation factor present; there is no change in the phase of the clock when seen from the reference frame in relative motion. In other words the clock is still synchronized with the clocks of the other reference frame. Let us consider now other clocks of the grid located at $x_1$, $x_2$, $x_3$, and so on. What happens to time intervals measured by these clocks when seen from a reference frame in relative motion? As mentioned they appear to be measuring an identical longer time interval. We have for each clock $t_2´ – t_1´$ = ($v/c^2$ x + $t_2$ – $v/c^2$ x – $t_1$)/sqrt (1 – $v^2/c^2$) = ($t_2$ – $t_1$)/sqrt(1 – $v^2/c^2$); the same result as for the clock located at the origin of the reference frame. But what about the phase of each clock located at x1, x2, x3, …, that are in synchrony in their reference frame (i.e. all the hands of the clocks are on the same mark), when seen from the reference frame in relative motion? We see that the clocks go at the same (apparently slower) rate but with a negative phase constant that increases with their distance to the origin. For example for the clocks locate at $x_1$ and $x_2$ we have t´ = ($v/c^2$ $x_1$ + t)/sqrt (1 – $v^2/c^2$) and t´ = ($v/c^2$ $x_2$ + t)/sqrt (1 – $v^2/c^2$); Here each $v/c^2$x can be seen as a factor that compensates the desynchronization of each clock (see, e.g., Bergman 1942, 38; Bohm 1965, 34; Iyer and Prabhu 2006; López-Ramos 2008). That is, in the reference frame in relative motion the clock located at $x_1$ will appear to be out of phase with the clock at the origin by – $v/c^2$ $x_1$, and the clock located at $x_2$ will appear to be out of phase with the clock at the origin by – $v/c^2$ $x_2$ (the same happens in relation to all other clocks). Thus, clocks that in their reference frame are synchronized marking, e.g., 12h, will appear from the perspective of a reference frame in relative motion to be out of phase. The clock at the origin will still be marking 12h, but the other clocks will be marking, e.g., 11.55, 11.50, 11.45 and so on. In this way *instead of talking about relativity (i.e the dependence on the reference frame) of simultaneity we might as well talk about the relativity (i.e. the dependence on the reference frame) of the synchronization of clocks.*[45]

---

44 A similar view is defended by Brown that considers that "the contraction and dilatation factors do not depend on, nor are an automatic result of, the relativity of simultaneity" (Brown 2005, 29, footnote 44).

Thinking in terms of relativity of clocks synchronization, we must now address its possible relevance regarding the notion of temporal simultaneity, i.e. of being present, sharing a common now. Let us consider four clones A, B, C, and D.  A, C, and D are located some distance apart in a shared reference frame (with clone A located at the origin, clone C located at – d, and clone D located at + d). Clone B is in relative motion in relation to the other clones. When clone B passes by A it synchronizes its clock to A's clock; let us say to $t_B$ = $t_A$ = 5 seconds.  Previously the clocks in A's reference frame had been synchronized. Thus to clone A, both C and D are taken to have the same phase as its clock (in this case and at this particular moment this means that the three clocks are marking $t_A$ = $t_C$ = $t_D$  = 5).[46] The synchronization procedure adopted in A's reference frame, that enables clone A to consider that clocks C and D are in phase with its clocks, from the perspective of clone B is in fact a desynchronization of the clocks in A's reference frame. To B the clock of C has a larger phase than the clock of A, and the clock of D has a smaller phase than the clock of A. This means that B will consider that the moment of D that for A corresponds to a phase of $t_D$ = 5 will correspond, e.g., to a phase of $t_D$ = 2, and while A regards a moment of C as $t_C$ = 5, B will take it to correspond, e.g., to $t_C$ = 8. This does not mean that C and D are present to A and are respectively in the future and past of B; what is happening is that A and B do not measure the same thing; only A is measuring the proper time of C and D by having first set/stipulate the relative phase of A, C, and D to zero. The apparent relativity of simultaneity is in fact a kinematical effect due to the relativity of synchronization of distant clocks.[47, 48] The confusion regarding the possible relevance of the relativity of synchronization regarding the notions of present, past and future is due to the fact that the question what time is it in C when B is marking $t_B$ is not a question about the spatial simultaneity (i.e. co-existence) of C and B, or about the temporal flow of each other. It is a question about the (eventual) synchronization of distant clocks, i.e. it is only meaningful when and to the point that a synchronization of distant clocks is made.

---

45 A similar view is given by Iyer and Prabhu that write "in this context, the principle of relativity of simultaneity can be restated as a principle of relativity of synchronicity as follows: spatially separated and synchronized clocks of any inertial frame appear asynchronous from any other inertial frame" (Iyer and Prabhu 2006, 3; see also Field 2006, 6)

46 As we have seen A does not have the information of the time keeping of the distant clocks of C and D instantaneously; it reconstructs it after, e.g., a signal arrives with this information.

47 A similar view is made by Field, that considers that "in standard special relativity theory, to date, this trivial difference in the synchronization convention [(due to the desynchronization factor – $v/c^2$ x)] given by the [Lorentz transformation] when x´ ≠ 0 has been interpreted as a real physical effect – 'relativity of simultaneity'" (Field 2006, 6; see also Logunov 1998, 33).

To the question how much time goes by in C when in B passes $\Delta t_B$ we can give an answer. In case of physical systems in relative motion it goes by the same duration, independently of having synchronized the clocks or not. The questions what time is it and what time goes by are different. Intuitively one tends to associate the idea of temporal co-presence (i.e. having the same now) to the first question. That is, to the idea that different distant clocks give the same time reading, but this depends on synchronization procedures and their limitation, i.e. the relativity of synchronization.

If we want we can rethink temporal co-presence as the sharing of the same temporal flow. As we have seen, physical systems in relative motion maintain their proper time (they go 'older' at the same pace). This is not the case when considering an accelerated physical system. However it is important not to confuse this to having different now-points. Even an accelerated physical system that, in relation to inertial systems, has its proper time 'slowed', is co-existent with the other physical systems (it does not slip into the past or future). To avoid any possible confusion, I prefer to maintain a view of temporal co-presence as the sharing of the same now-point that is independent of the temporal flow of the physical systems proper times, i.e. just in terms of the spatial co-existence/co-presence of the different physical systems.

We arrive at a view of present different from Dieks's. As we have seen, Dieks proposes a now-point per worldline view. Here it is being proposed the same now-point to every worldline in terms of a spatial co-existence of the physical systems. As it is made clear, in particular by considering accelerated physical systems, *the now is defined not by the temporal flux of each physical system, but by a shared spatiality*.

This might seem to be a return to a Newtonian-like view of time; an absolute time that flows independently of the physical systems but shared by all, which is beyond space but 'sensed' in every point of the absolute Newtonian space. That is not the case. In this view, like with Dieks's, one does not regard time independently of physical systems; on the contrary, time is thought in terms of proper time. This means that independently of adopting Dieks's view or the one being proposed here we face a clear change in the conception of time from the Newtonian notion to the relativistic one. Time is proper time; it is local, or better, it is embodied in the physical systems.

---

48 Let us consider two thunders that strike at the spatial locations of C and D at the moment $t_A = t_C = t_D = 5$. B will, besides attributing a different location of C and D in its reference frame, measure a time interval between the two events different from zero. For B, $t_C' - t_D' = \beta(x_D - x_C)v/c^2$. The key element here is not $\beta$, which is related to the kinematical time dilatation effect of improper time measurements; it is in the desynchronicity factor for each clock, i.e. $-xv/c^2$. From our previous considerations one cannot take this result as implying that while for A, the thunders are spatially co-existent, the same is not true to B. C and D are taken to have the same 'age' as A and B (i.e. an identical proper time, since they were never under acceleration); also they are supposed to have their 'internal clock' phase synchronized with that of A's. This is an undisputed 'startup condition' for setting forward a reference frame: (1) one can consider the equality of proper times of identical clones located face to face as an empirical result; (2) the equality of proper times of distant clones or identical clocks in relative motion follows from the principle of relativity, being implicit the so-called boostability assumption; (3) one accepts that after synchronization there is an actual physical meaning for saying that two events occur at the same time at distant locations (and this might be debatable under the conventionality thesis; see, e.g., Brown 2005, 19-22 and 95-98). Accepting these points, in particular 3 (and this is made in the usual interpretation in terms of the so-called relativity of simultaneity), if let us say clones C and D are just 12 h, 05 min, and 3 s old and two thunder strike 'now' near C and D, then one can say that these two events co-exist spatially; and because B co-exists with A, C and D, it also co-exists with the two thunder striking, independently of, due to the relativity of synchronization, it will attribute to each thunder strike (and to each clone C and D age) a different phase (i.e. a different temporal value). To put in other words, if it is physically meaningful within the theory of relativity to say that distant identical physical systems having the same proper time can be set in phase (meaning by this that they have the same internal clock that begun running, for example when the two physical systems where side by side, and that they are still synchronized when located at a distance) then it is also meaningful to say regarding two events (that can be seen/idealized as points of these physical systems worldliness) that these events are actually simultaneous, or in my phrasing that these events are spatially co-existent with all physical systems, i.e. they do not have different now-points depending on the reference frame.

9 The completion of Galilean space-time: time as proper time

From the previous section one might think that we have a clear transformation of our conception of time from the Newtonian to the Einsteinian framework. That does not have to be the case. As we have seen the difference between the Newtonian and the Galilean space-time is in the breakdown of the notion of absolute space. It simply is not necessary in the Newtonian kinematics or dynamics. As we have seen, historically the concept of absolute space had been criticized previous to Einstein's work. However it was only after the insights brought by Einstein that a Galilean space-time was actually formulated. The view defended here is that this formulation is not yet completed and that one must follow the insight about time made possible by the theory of relativity one step further. This mean to let go of absolute time, which is still apparently present in the Galilean space-time, and to learn from the theory of relativity to think about time in terms of proper times.[49] In fact we can consider that the Galilean space-time can be described by using the proper times of physical systems exactly by the same argument used in the theory of relativity, in which we can see the notion of proper time of a clock as more fundamental than the notion of coordinated time. What creates the 'illusion' of a universal independent time is the fact that within the Galilean framework there is direct action-at-a-distance, or what is the same the signals/interactions between physical systems are taken to have infinite velocity (i.e. they are instantaneous). This implies that all clocks of a reference frame (or between reference frames) can be synchronized instantaneously.[50] This gives the impression that time is 'out there' on its own in space; however what is going on is, like in the theory of relativity, a spreading of synchrony between physical system locates at a distance (Brown 2005, 19-20; see also Kapuścik 1986). With this completion of a Galilean space-time without absolute space and without absolute time (in the pre-relativistic sense) we can see the crucial aspect of time in the theory of relativity as proper time as shared also by the Galilean space-time.[51]

10 Conclusion

In the previous section I have tried to close the circle regarding a view of Galilean space-time that enables a greater continuity between the notion of space and time previous to the theory of relativity and the ones deployed with the theory. As it is well-known, mathematically, there is no fundamental difference between the theory of relativity and Lorentz's electron theory. As Darrigol called the attention to, "the empirical equivalence of the two theories simply results from the fact that any valid reasoning of Einstein's theory can be translated into a valid reasoning [in Lorentz's electron theory]" (Darrigol 2006, 19 footnote 28). In fact we can see Einstein's contribution essentially as the deployment of a new conceptual system within an already existent mathematical structure (see, e.g., Merleau-Ponty 1974, 113-6). By proposing a reevaluation of the concepts of space and time in the theory of relativity in terms of a strong continuity with previous theories is one downplaying the importance of Einstein's contribution? No. Let us see why, by giving an explicit answer to the title-question of this work. If we give an answer from an historical perspective, in fact Einstein made possible the development of relativistic notions of space and time, which did not exist before him.[52] As mentioned, the development of Galilean relativistic notions of space and time is posterior to the theory of relativity; and so in simple terms if one asks did the concepts of space and time change that much with the 1905 theory of relativity? The answer from an historical perspective is yes. If now one forgets about the particular historical contingency of the development of the theory of relativity in the context of an existing electron theory, and considers, from the insight provided by Einstein, Newtonian mechanics by taking into account the Galilean principle relativity, one can make a conceptual 'cleanup' of Newtonian mechanics as it was actually done after, and proposed in part before, Einstein. In doing this one arrives at the conclusion that Newton's absolute space and absolute time are not a necessary presupposition of Newtonian mechanics. In this way one arrives at a Galilean notion of space and time, i.e. a notion of space and of time that takes into account the Galilean principle of relativity. If, following Brown, one makes a similar fable about an Albert Keinstein (Brown 2005, 33-40), which we imagine that around 1705 arrives at a Galilean notion of space and time, and now asks if the theory of relativity brought a throughout change regarding the conception of space and time in physics, one can answer, in a very simplified way, that it did not. That is, from an an-historical perspective by putting side by side the theory of relativity and Newtonian mechanics with a Galilean space-time one can say that there is an important continuity in the concept of space in terms of an homogeneous spatially and of time in terms of the notion of proper time.

---

49 I think one can read Brown's considerations regarding the inexistence of a time dilatation and the conventionality in distant simultaneity within Newtonian mechanics in terms of an implicit use of proper time instead of absolute independent time (see Brown 2005, 19).
50 Brown gives a different view in terms of a standard convention in which time is spread through space "in inertial frames in such a way that actions-at-a-distance like gravity are instantaneous" (Brown 2005, 20).
51 From a different perspective, Brown considers that, in relation to aspects shared with 'Newtonian time' related to the so-called relativity of simultaneity, "the common claim that Einstein revolutionized the notion of time seems to me to be overstated" (Brown 2005, 20).

References

Arthur, R. T. W. (2010). Minkowski's proper time and the status of the clock hypothesis, in V. Petkov (ed.), Space, time and spacetime. Fundamental theories of physics, Vol. 167, part 2, pp. 159-179. Heidelberg: Springer.
Bergmann, P. G. (1976 [1942]). Introduction to the theory of relativity. New York: Dover publications.
Bohm, D. (1996 [1965]). The special theory of relativity. London and New York: Routledge.
Brown, H. R. (2005). Physical relativity: space-time structure from a dynamical perspective. Oxford: Clarendon Press.
Brown, H. R. and Pooley, O. (2006). Minkowski space-time: a glorious non-entity, in D. Dieks (ed.), The ontology of spacetime, pp 67-89. Amsterdam: Elsevier.
Callahan, J. J. (2000). The geometry of spacetime: an introduction to special and general relativity. New York: Springer.
Clavelin, M. (1996 [1968]). La philosophie naturelle de Galilée. Paris : Albin Michel.
Cresser, J. D. (2003). The special theory of relativity: lecture notes. Department of physics, Macquarie University.
Darrigol, O. (2000). Electrodynamics from Ampere to Einstein. Oxford: Oxford University Press.
Darrigol, O. (2006). The genesis of the theory of relativity, in T. Damour, O. Darrigol, B. Duplantier, and V. Rivasseau (eds.) Einstein, 1905-2005, pp 1-31. Basel: Birkhäuser.
Darrigol, O. (2007). A Helmholtzian approach to space and time. Studies in History and Philosophy of Science, 38, 528-542.
Dieks, D. (1988). Special relativity and the flow of time. Philosophy of science, 55, 456-460.
Dieks, D. (2006). Becoming, relativity and locality, in D. Dieks (ed.), The ontology of spacetime,

---

[52] Let us not forget that to Lorentz the so-called local time was just a mathematical artifact to help in calculations (see, e.g. Lorentz 1916, 57-8 and 187-9; Darrigol 2006, 11). Poincaré did notice that the local time had operational meaning since actually this is the time measured by an observer in absolute motion in relation to the ether; However Poincaré still considered a sort of absolute time ('le temps réel'), the time measured by an observer in absolute rest in relation to the ether (see, e.g., Poincaré 1913, 43-6; Darrigol 2006, 17-9). It was Einstein who for the first time presented a truly relativistic notion of time in which physical time has to be defined and 'spread' within an inertial reference frame, and there is no preferred reference frame (see, e.g., Einstein 1905; Paty 1993, 148-52).

Vol. 1. Amsterdam: Elsevier.
DiSalle, R. (2009). Space and time: inertial frames. Stanford Encyclopedia of Philosophy, Plato.Stanford.edu/entries/spacetie-frames/ (accessed on November 2011)
Dutta, M., Mukherjee, T. K., and Sen, M. K. (1970). On linearity in the special theory of relativity. International Journal of Theoretical Physics, 3, 85-91.
Earman, J. (1989). World enough and space-time. Cambridge: MIT Press.
Einstein, A. (1989 [1905]). On the electrodynamics of moving bodies. In, The collected papers of Albert Einstein (English translation), Vol. 2, pp. 140-171. Princeton: Princeton University Press.
Einstein, A. (1989 [1907]). On the relativity principle and the conclusions drawn from it. In, The collected papers of Albert Einstein (English translation), Vol. 2, pp. 252-311. Princeton: Princeton University Press.
Einstein, A. (1993 [1910]). The principle of relativity and its consequences in modern physics. In, The collected papers of Albert Einstein (English translation), Vol. 3, pp. 117-142. Princeton: Princeton University Press.
Einstein, A. (1993 [1911]) The theory of relativity. In, The collected papers of Albert Einstein (English translation), Vol. 3, pp. 340-350. Princeton: Princeton University Press.
Einstein, A. (1996 [1912-1914]). Manuscript on special relativity. In, The collected papers of Albert Einstein (English translation), Vol. 4, pp. 3-88. Princeton: Princeton University Press.
Einstein, A. (1997 [1914]). On the principle of relativity. In, The collected papers of Albert Einstein (English translation), Vol. 6, pp. 3-5. Princeton: Princeton University Press.
Einstein, A. (1996 [1915]). Theory of relativity. In, The collected papers of Albert Einstein (English translation), Vol. 4, pp. 246-263. Princeton: Princeton University Press.
Einstein, A. (2002 [1920]). Ether and the theory of relativity. In, The collected papers of Albert Einstein (English translation), Vol. 7, pp.160-182. Princeton: Princeton University Press.
Field, J. H. (2006). Clocks rates, clock settings and the physics of the space-time Lorentz transformation. ArXiv: physics/0606101v4.
Friedman, M (1983). Foundations of space-time theories. Princeton: Princeton University Press.
Havas P. (1964). Four-dimensional formulation of Newtonian mechanics and their relation to the special theory of relativity. Review of Modern Physics, 36, 938-965.
Iyer, C., and Prabhu, G. M. (2006). Constructive derivation of asynchronicity. Department of Computer Science, Iowa State University.
Jammer, M. (1993). Concepts of space. New York: Dover publications.
Janis, A. (2010). Conventionality of simultaneity. Stanford Encyclopedia of Philosophy, plato.Stanford.edu/entries/spacetime-convensimul/ (accessed on November 2011).
Janssen, M (1995). A Comparison Between Lorentz's Ether Theory and Special Relativity in the Light of the Experiments of Trouton and Noble. Ph.D. Dissertation, University of Pittsburgh.
Kapuścik, E. (1986). On the physical meaning of the Galilean space-time coordinates. Acta Physica Polonica, 7, 569-575.
Kosto, L. (2000). Einstein and the ether. Montreal: Apeiron.
Koyré, A. (1958 [1957]). From the closed world to the infinite universe. New York: Harper.
Logunov, A. A. (1998). Curso de teoría de la relatividad y de la gravitación. Moscú: Editorial URSS.
López-Ramos, A. (2008). Time desynchronization and Ehrenfest paradox. ArXiv: 0803.2036v1.
Lorentz, H. A. (1916). The theory of electrons and its applications to the phenomena of light and radiant heat. New York: G. E. Stechert & Co.
Merleau-Ponty, J. (1974). Leçons sur la genèse des théories physiques. Paris : Libraire Philosophique J. Vrin.
Minkowski, H. (1908). Space and time. In, The principle of relativity, pp. 75-91. New York: Dover publications.
Misner, C. W., Thorne, K. S., and Wheeler, J. A. (1973). Gravitation. San Francisco: Freeman.
Norton, J. D. (1993). General covariance and the foundations of general relativity: eight decades of dispute. Reports on Progress in Physics 56, 791-858.

Norton, J. D. (2008). Why constructive relativity fails. The British Journal for the Philosophy of Science, 59, 821-834.
Pauli, W. (1958 [1981]). Theory of relativity. New York: Dover publications.
Paty, M. (1993). Einstein philosophe. Paris: Presses Universitaires de France.
Poincaré, H. (1898). La mesure du temps. Revue Métaphysique et de Morale, 6, 1-13.
Poincaré, H. (1913). La dynamique de l'électron. Supplément aux Annales des Postes Télégraphes et Téléphones.
Reichenbach, H (1957 [1927]). The philosophy of space and time. New York: Dover publications.
Saint-Ours, A. (2008). Time and Relation in Relativity and Quantum Gravity: From Time to Processes. In D. Dieks (ed.), The Ontology of Spacetime II. Amsterdam: Elsevier.
Savitt, S. (2011). Time in the special theory of relativity. In C. Callender (ed.), The Oxford Handbook of Philosophy of Time. Oxford: Oxford University Press.
Schutz, B. F. (1985). A first course in general relativity. Cambridge: Cambridge University Press.
Smith, J. A. (1995 [1965]). Introduction to special relativity. New York: Dover publications.
Stephenson, G., and Kilmister, C. W. (1958). Special relativity for physicists. London: Longmans.
Taylor, E. F., and Wheeler, J. A. (1965). Spacetime physics. San Francisco: Freeman.
Tolman, R. C. (1987[1934]). Relativity thermodynamics and cosmology. New York: Dover publications.
Torretti, R. (1996). Relativity and geometry. New York: Dover publications.
Torretti, R. (2007). De Eudoxo a Newton. Santiago, Chile: Ediciones Universidad Diego Portales
Wald, R. M. (1984). General relativity. Chicago: University of Chicago Press.
Weyl, H. (1952 [1921]). Space, time, matter. New York: Dover publications.
Weyl, H (1949). Philosophy of mathematics and natural science. Princeton: Princeton University Press.
Whittaker, E. (1960 [1953]). A history of the theories of aether and electricity, Vol. 2. New York: Harper.
Zhang, Y. Z. (1997). Special relativity and its experimental foundations. Singapore: World Scientific.